\documentstyle[12pt]{article}

\def\non{\nonumber}
\newcommand{\reff}[1]{(\ref{#1})}
\def\al{\alpha}

\def\D{{\cal D}}
\def\L{{\cal L}}
\def\de{\delta}

\author{S. M. Klishevich, Yu. M. Zinoviev
        \thanks{E-mail address: ZINOVIEV@MX.IHEP.SU} \\
        {\it Institute for High Energy Physics} \\
        {\it Protvino, Moscow Region, 142284, Russia}}
\title{On electromagnetic interaction of massive spin-2 particle}
\date{June 1997}

\begin{document}
\maketitle

\begin{abstract}
In this work we construct a gauge invariant description of free
massive particle with an arbitrary integer spin. Such description
allows one to investigate the problem of consistent interactions for
massive high spin particles using the requirement that the whole
interaction Lagrangian must be gauge invariant. As a first example of
such approach, we consider the case of electromagnetic interaction of
massive spin-2 particle: a linear approximation in a case of the
arbitrary field and a full theory for the homogeneous electromagnetic
field in the space-time of any dimensionality.
\end{abstract}

\newpage

\section*{Introduction}

   The interactions of the fields with spins $s \le 2$ are quite
well studied, while about those of higher spins we know much less. In
all investigations of high spin fields the Poincare group plays a
special role. It not only determines the most important kinematical
properties of the particles, but also fixes, to a great extent, 
their interactions. Namely, the covariant description of particles
with spins $s \ge 1$ necessarily requires the presence of the
invariance under the gauge transformations and one has to keep this
invariance at the switching on the interaction. Using such a
requirement, in the investigations of gauge invariant interactions of
spin-1 particles one unambiguously comes to the Yang-Mills theory,
while starting from free spin-2 field and switching on the
interaction, one can reproduce the usual gravity theory, see, for
example, \cite{Ogiev,Fronsdal-2}. In the same way, the requirement of
gauge invariance for the spin-2 and spin-3/2 particles interactions
leads to supergravity \cite{Milton}.

   It is known, that the minimal interactions of any massless spin
$s>2$ particles with vector fields (abelian or non-abelian) turns out
to be inconsistent\footnote{This result has been formulated as a
theorem in \cite{Zinoviev-2}} . Analogously, all the attempts to
switch on the gravitational interaction for the massless spin $s>2$
particles lead to the inconsistencies. It seems, that there are no
consistent theories for the interactions of such particles (in flat
space). Till now, the only consistent theory with the high spin
particles interactions is the superstring theory. The massless sector
of this theory is restricted with the states of spins $s \le 2$,
while the massive sectors contain the particles of arbitrary spins.
All these particles with all spins enter the string interaction, but
it would be interesting to know the peculiarity of the interaction
for concrete particle with definite spin. Unfortunately, it is hard
to extract the information on the gauge symmetries or interactions
for particular state in the string spectrum and no essential results
in this direction were achieved.

   Note, that there exist consistent theories of the interaction of
high spin massless particles in the space of constant nonzero
curvature, which were studied in \cite{Vasiliev}. In principle,
starting from such theories one could obtain more physically
interesting case of massive particles in the flat space. For that,
one has to solve the problem of spontaneous gauge symmetry breaking
in these theories, but it is a very complicated task and it has not
been solved up to now.

  The description for the free massive particles with arbitrary spin
was given in a well known paper \cite{Sing-Hagen}. But the Lagrangian
for the massive particle in this approach, unlike the one for the
massless particles, does not possess gauge invariance. So, for the
construction of the consistent interaction theory one has to
introduce, besides the usual requirement of the Lorentz invariance,
some additional considerations. For example, in paper \cite{FPT-92},
where the electromagnetic interactions of high spin particles were
investigated, the requirement that the tree level amplitudes must
have smooth massless limit with fixed charge was used. In turn, in
papers \cite{P-93,CDP-94} it was proposed to construct gravitational
interaction for high spin particles using a tree level unitarity as
such additional requirement.

   For the spin-1 and spin-3/2 particles there is a well known
mechanism --- mechanism of spontaneous symmetry breaking. In this,
one starts with the consistent massless theory and then introduces
masses. But as we have already mentioned, it seems that there are no
consistent theories for massless spin $s > 2$ particles (in flat
space). The main property of spontaneous symmetry breaking in both
cases is the possibility to have gauge invariant description of
massive spin-1 or spin-3/2 particles due to the introduction of
Goldstone fields with inhomogeneous transformation laws. In paper
\cite{Zinoviev-1} the method for the construction of massive high
spin gauge fields interactions based on the gauge invariant
description of such particles was proposed. It allows one to keep the
principle of gauge invariance as the fundamental one for the
construction of the interaction of massive particles in the same way
as it was used for the massless ones. In paper \cite{Zinoviev-1} the
gauge invariant description for massive particles with spins up to
$s=3$ was given and as a demonstration the simplest possible case of
spontaneous symmetry breaking in the Yang-Mills theory with $SO(3)$
group was considered. It was shown that the approach proposed allows
one to reproduce two well known possibilities: non-linear
$\sigma$-model, see e.g. \cite{Slavnov}, and, with the help of the
introduction of additional scalar field, the usual model of
spontaneous symmetry breaking with the doublet of Higgs fields.

  In this paper, in the next Section, using the method proposed in
\cite{Zinoviev-1} we construct a gauge invariant description of free
massive particles with arbitrary integer spin.

  Among all the interactions, the electromagnetic interaction is one
of the most well studied and for a long time it serves as a polygon
for the investigation of different models. Thus, it seems natural to
start with the investigations of electromagnetic interactions for the
massive high spin particles. In Section 2 we begin with the linear
approximation for the e/m interaction of massive spin-2 particle. The
construction of the linear approximation is a very important step
in studying any theory of high spin interactions, because this
approximation does not depend on the presence of any other fields in
the theory, while all higher approximations are heavily model
dependent. Thus, the linear approximation turns out to be universal
for any theory (with the given number of derivatives, of course).

   The next to linear approximation requires a lot of calculations
even in the minimal model without the introduction of any additional
fields. The calculations become much easier, then one considers the
case of homogeneous electromagnetic field. Recently, in
\cite{Argyres} the model describing massive spin-2 particle moving in
homogeneous e/m field has been constructed. But the authors of
\cite{Argyres} started from the bosonic string, so, their results
hold for the $d=26$ dimension only. In Section 3, using constructive
approach, we obtain an analogous result, but for the space of
arbitrary dimension.

\section{Gauge Invariant Description of Free Massive Particle with
 Arbitrary Integer Spin}
\label{free-spin}

\subsection{Massless Particle Lagrangian}

For the description of massless spin-S particle we will use the
formalism proposed in \cite{Fronsdal-1}.
  Let us consider a symmetric tensor field of rank s --- 
$ \Phi^s=\Phi^{\mu_1\ldots\mu_s} $, where Greek indices take the
following values: $\mu,\nu,\ldots = 0,\ldots\,3$ and require this
field to be double traceless
 $$ \widetilde{\widetilde{\Phi}}=Sp\,(\,Sp\,\Phi^s\,)\,=\,0 ,$$
here  $ \widetilde{\Phi}\stackrel{def}{=}Sp\,\Phi^s $,
$\widetilde{\widetilde{\Phi}}=Sp\,Sp\,\Phi^s $ and so on, $Sp$ is a
contraction of two indices by metric tensor.

The gauge transformation for $\Phi^s$ is taken as
\begin{equation}
\label{D0}
\delta_o \Phi^s = \frac{1}{s} \cdot
\left\{\partial\Lambda^{s-1}\right\}_{s}
\end{equation}
where $\{\ldots\}_{s.}$ means symmetrization over all the indices
(without normalization) and $\Lambda^{s-1}$ is symmetric traceless
tensor field of rank $s-1$.

Let us write the most general quadratic Lagrangian with two
derivatives for field $\Phi^s $. Using the freedom in the choice of
field normalization, we can fix the coefficient at
$(\partial_\mu\Phi^s)^2 $ so that this term has the standard form
\begin{eqnarray}
  \L_o &\!=\!& \frac{ \left(-1\right)^s }{2}(\partial_\mu\Phi^s)
    \cdot(\partial_\mu\Phi^s)
   +a_1(\partial\cdot\Phi^s)\cdot(\partial\cdot\Phi^s)
   + a_2(\partial_\mu\widetilde{\Phi}^s)\cdot
    (\partial_\mu\widetilde{\Phi}^s)
\non\\ &&{}
   +a_3(\partial\cdot\Phi^s)\cdot(%
\stackrel{\leftarrow}{\partial}\widetilde{\Phi}^s)+a_4(\partial
\cdot\widetilde{\Phi}^s)\cdot(\partial\cdot\widetilde{\Phi}^s),
\label{eq:2}
\end{eqnarray}
  where $(\partial\cdot\Phi^s)\stackrel{def}{=}\partial_\mu
\Phi^{\mu\mu_2\ldots\mu_s}$, $(\partial\cdot\partial\cdot\Phi^s)
\stackrel{def}{=}\partial_\mu\partial_\nu\Phi^{\mu\nu\mu_3\ldots\mu_s
}$, and the point denotes contraction of all the indices between
tensor objects, for example, $\Phi^s \cdot \Phi^s \stackrel{def}{=}
\Phi_{\mu_1\ldots\mu_s} \Phi^{\mu_1\ldots\mu_s}$.

Calculating a variation\footnote{We neglect the terms proportional to
a total derivative.} \  of the Lagrangian under
transformation~\reff{D0} one obtains
\begin{eqnarray}\non
  \delta_o\L_o &\!\!=\!\!&
\left(\!\left(-1\right)^s+\frac{2a_1}{s}\right)\!
                     \partial^2(\partial\cdot\Phi^s)\Lambda^{s-1}
    +\frac{2}{s}\left(\left(s-1\right)a_1+a_3\right)\!
    \partial(\partial\cdot\partial\cdot\Phi^s)\Lambda^{s-1}  
\\\non &&{}
  +\frac{2}{s}\left(a_3+2a_2\right)\!\partial^2
(\partial\widetilde{\Phi}^s)
   \Lambda^{s-1}+\frac{1}{s}\left(\left(s-2\right)a_3+4a_4\right)\!
    \partial\partial(\partial\cdot\widetilde{\Phi}^s)\Lambda^{s-1}.
\end{eqnarray}
From the condition of the invariance of Lagrangian~\reff{eq:2}
under the gauge transformation~\reff{D0}, one get simple
equations on arbitrary coefficients in the Lagrangian. Solving these
equations, we obtain a final form for the Lagrangian of a free
massless spin-$s$ particle
\begin{eqnarray*}
  \L_o&\!=\!&(-1)^s\biggl\{\frac{1}{2}(\partial_\mu\Phi^s)
(\partial_\mu\Phi^s)
      -\frac{s}{2}(\partial\cdot\Phi^s)(\partial\cdot\Phi^s)
      -\frac{s(s-1)}{4}(\partial_\mu\widetilde{\Phi}^s)
                       (\partial_\mu\widetilde{\Phi}^s)  
\\ &&{}
    -\frac{s(s-1)}{2}(\partial\cdot\partial\cdot\Phi^s)\widetilde
{\Phi}^s
      -\frac{1}{8}s(s-1)(s-2)(\partial\cdot\widetilde{\Phi}^s)
      (\partial\cdot\widetilde{\Phi}^s)\biggr\}
\end{eqnarray*}

One can get the same result from the Lagrangian for massive particle 
proposed in \cite{Sing-Hagen}, if one let $m\to0$ and make some
field redefinition.

\subsection{Lagrangian for Massive Particle}

As is well known, it is not enough to have only one field $\Phi^s$
for the correct description of massive spin-$s$ particle.
One has to introduce some lower spin auxiliary fields. For instance,
in \cite{Sing-Hagen} the authors considered a set of a symmetric
traceless fields of ranks $ 0,1,\ldots,{s-2},s$ and demanded that the
auxiliary fields with spins $0,1,\ldots,s-2$ would vanish on the
equations of motion, while the equations for the field $\Phi^s$ had
the usual form
$$
 (\partial^2+m^2)\Phi^{\mu_1\ldots\mu_s}=0,
 \qquad\partial_\mu\Phi^{\mu\mu_2\ldots\mu_s}=0.
$$
But the resulting Lagrangian of massive particle was not gauge
invariant, moreover, in the massless limit $m\to0$ the number of
physical degrees of freedom changes from $2s+1$ to $2$.

Below we construct an alternative gauge invariant formalism for the
description of massive spin-$s$ particles, whose massless limit gives
a sum of the massless particles of spins $s,s-1,...0$. This approach,
which was offered in \cite{Zinoviev-1}, is based on the possibility
to have gauge invariance for the massive free particles due to the
introduction of additional fields, corresponding to all lower spins.
As it has already been mentioned in the Introduction, such approach
allows one to investigate the interactions of massive particles with
arbitrary spins\footnote{In this paper we will deal with the integer
spin particles only.} \ in the same way as the ones of the massless
particles.

We will start with the Lagrangian describing the sum of free massless 
particles with spins $0,1,...,s$ and we will add quadratic terms
proportional to $m$ (with one derivative) and to $m^2$ (without
derivatives) keeping all the gauge invariances of initial massless
fields. As a result, one will get the description of free massive
spin-$s$ particle, where, by construction, in the massless limit
$m\to0$ the Lagrangian will break into the sum of Lagrangians
corresponding to massless particles with spins $0,1,\ldots,s$, so
that the number of physical degrees of freedom remains to be the same
and the gauge invariance is present both on the massive as well as
on the massless level.

Following the program described above, we introduce a set of
symmetric double traceless fields
$\{\Phi^0,\Phi^1,\ldots,\Phi^{s-1},\Phi^s\}$
and, correspondingly, a set of traceless gauge parameters 
$\{\Lambda^0,\,\Lambda^1,\ldots,\Lambda^{s-2},\Lambda^{s-1}\}$.
Let us write the most general form of gauge transformations for
fields $\{\Phi^0,\Phi^1,\ldots, \Phi^{s-1}\}$ in the presence of a
dimensional parameter $m$:
\begin{equation}
\label{tr}
\delta_o\Phi^k=\frac{1}{k}\left\{\partial\Lambda^{k-1}\right\}_{s.}
+m\left[c_1(k)\Lambda^k+c_2(k)\left\{g^2\Lambda^{k-2}\right\}_{s.}
\right],
\end{equation}
where the first term is absent at $k=0$ and the third one at $k<2$,
while $g^2$ is a metric tensor.

Now we write the most general quadratic Lagrangian with, at most, two
derivatives:
\begin{eqnarray}
   \label{Lmo}
   \L_o&=&\sum_{k=0}^{s}\biggl\{\L_o(k)+m[a_1(k)\Phi^{k-1}
(\partial\cdot\Phi^k)
      +a_2(k)\widetilde{\Phi}^k(\partial\cdot\Phi^{k-1})  \nonumber\\
&&    +a_3(k)(\partial\cdot\widetilde{\Phi}^k)\Phi^{k-3}
      +a_4(k)(\partial\cdot\widetilde{\Phi}^k)\widetilde{\Phi}^{k-1}]
      \nonumber\\
&&    +m^2[b_1(k){(\Phi^k)}^2+b_2(k){({\widetilde{\Phi}}^k)}^2
      +b_3(k){\widetilde\Phi}^k\Phi^{k-2}]\biggr\},
\end{eqnarray}
where $\L_o(k)$ is the Lagrangian for a massless spin-$k$ particle.
Requiring that Lagrangian~\reff{Lmo} be invariant under
transformations~\reff{tr}, one obtains the algebraic system of
homogeneous equations for the arbitrary coefficients
\begin{eqnarray*}
&&   a_3(k)=0,\\
&&   (-1)^kc_1(k)-\frac{a_1(k+1)}{k+1}=0,\\
&&   (-1)^{k+1}kc(k)-\frac{1}{k+1}(ka_1(k+1)-2a_2(k+1))=0,\\
&&   \frac{(-1)^k}{2}(k-1)kc_1(k)-\frac{2}{k+1}a_4(k+1)=0,\\
&&   (-1)^{k+1}(k-1)^2kc_2(k)-\frac{a_2(k)}{k-1}=0,\\
&&   (-1)^kk(k-1)kc_2(k)+a_1(k)=0,\\
&&   \frac{(-1)^{k+1}}{2}(k-2)(k-1)^2kc_2(k)-\frac{k-2}{k-1}a_2(k)
    +\frac{2}{k-1}a_4(k)=0,\\
&&   a_1(k+1)c_1(k)-2b_1(k+1)=0,\\
&&   a_2(k)c_1(k)+\frac{2}{k+1}b_3(k+1)=0,\\
&&   a_2(k+1)c_1(k)+\frac{4}{k+1}b_2(k+1)=0,\\
&&   c_2(k)\biggl[\frac{(k-1)k}{2}a_1(k+1)-(k-1)a_2(k+1)
      +2ka_4(k+1)\biggr]  \\
&&{} -b_3(k+1)=0,
\end{eqnarray*}
which allows one to express all the coefficients through the
parameters $c_1(k)$:
\begin{eqnarray}
   c_2(k)&=&\frac{c_1(k-1)}{(k-1)^2},  \nonumber\\
   a_1(k)&=&(-1)^{k+1}k\,c_1(k-1),  \nonumber\\
   a_2(k)&=&(-1)^{k+1}(k-1)k\,c_1(k-1),  \nonumber\\
   a_3(k)&=&0,   \nonumber\\
   a_4(k)&=&\frac{(-1)^{k+1}}{4}(k-2)(k-1)k\,c_1(k-1),  \label{k} \\
   b_1(k)&=&\frac{(-1)^{k+1}}{2}\left(k{c_1}^2(k-1)-
        \frac{(k+1)(2k+1)}{k}{c_1}^2(k)\right), \nonumber\\
   b_2(k)&=&\frac{(-1)^{k+1}}{4}\left(\frac{(k^2-1)(k+2)}{2}
        {c_1}^2(k)- (k-1)k^2{c_1}^2(k-1)\right),   \nonumber\\
   b_3(k)&=&\frac{(-1)^{k+1}}{2}(k-1)kc_1(k-2)c_1(k-1), \nonumber
\end{eqnarray}
in this, we have the recurrent relation for the $c_1(k)$:
$$
{c_1}^2(k-1)-\frac{(k+1)(2k+1)}{k^2}{c_1}^2(k)+
\frac{(k+2)^2}{k(k+1)}{c_1}^2(k+1)=0.
$$
In order to solve this relation unambiguously, we must fix one of
the parameters $c_1(k)$. Let us require that the dimensional
parameter $m$ be a mass of the particle, i.e. let us put
$b_1(s)=\frac{(-1)^{s+1}}{2}$. Then solving the recurrent relation,
we obtain
\begin{equation}
   c_1(k)=\frac{1}{k+1}\sqrt{\frac{(s-k)(s+k+1)}{2}}. 
\label{c}
\end{equation}
Putting \reff{c} in \reff{k} yields the final result.
It is not difficult to calculate all the coefficients in~\reff{Lmo}
for each concrete case, but the general formula for an arbitrary spin
is rather cumbersome, therefore, we will not write it here.

Thus, we have constructed the Lagrangian for a free massive spin-$s$
particle which is invariant under the gauge
transformations~\reff{tr}, in this, the Lagrangian has the correct
massless limit. Let us note that one can, in principle, rewrite all
the formulas in terms of the unconstrained tensors. Namely, one can
join double traceless field $\Phi^4$ with $\Phi^0$ into one
unconstrained rank-4 tensor, then $\Phi^5$ with $\Phi^1$ and so on.
Analogously, combining traceless gauge parameter $\Lambda^2$ with
$\Lambda^0$, one gets unconstrained rank-2 tensor and so on. It is
easy to see that one ends with just four fields $\Phi^s$,
$\Phi^{s-1}$, $\Phi^{s-2}$ and $\Phi^{s-3}$ and two gauge parameters
$\Lambda^{s-1}$ and $\Lambda^{s-2}$, exactly as in \cite{Pashnev}.

From the description of massive spin-$s$ particle given above one can
reproduce the well-known result obtained in \cite{Sing-Hagen}. In
order to show this, we represent the symmetric double traceless
tensor field $\Phi^k$ as $$ \Phi^k = {\Phi'}^k + \frac{1}{2k}
\{g^2\widetilde{\Phi}^k\}, \qquad k=0,1,\ldots,s-1\ ,$$
where ${\Phi'}^k$ is a symmetric traceless tensor field. Using the
gauge transformations we can always exclude the fields
$\{{\Phi'}^0,{\Phi'}^1, \ldots,{\Phi'}^{s-1}\}$ by some choice of
parameters $\Lambda^k$. Redefining ${\widetilde\Phi}^k=\Phi^{k-2}$,
we obtain the set of the symmetric traceless tensor fields
$\{\Phi^s,\Phi^{s-2},\ldots,\Phi^0\}$, which is equivalent to the one
in~\cite{Sing-Hagen}. In this, the Lagrangian has the following form:
\begin{eqnarray}
  \L_o&=&(-1)^s\biggl\{\frac{1}{2}{(\partial_\mu\Phi^s)}^2-
\frac{s}{2}    {(\partial\cdot\Phi^s)}^2
-\frac{(s-1)^2}{2}(\partial\cdot\partial
     \cdot\Phi^s)\Phi^{s-2}  \nonumber\\
&&   -\frac{(s-1)^2(2s-1)}{8s}(\partial_\mu\Phi^{s-2})^2-
    \frac{(s-1)^2(s-2)^2}{8s}{(\partial\cdot\Phi^{s-2})}^2
\nonumber\\
&& - \biggl[\frac{1}{2}(\Phi^s)^2-\frac{(s-1)(2s-1)}{8}(\Phi^{s-2})^2
    \biggr]+\sum_{q=3}^s\biggl(A_1^{s-q}(\partial_\mu\Phi^{s-q})^2
   \nonumber\\
&& + A_2^{s-q}(\partial\cdot\Phi^{s-q})^2+
         mB^{s-q+1}\Phi^{s-q}(\partial\cdot\Phi^{s-q+1})+
              m^2C^{s-q}(\Phi^{s-q})^2\biggr)\biggr\}, \nonumber
\end{eqnarray}
where
\begin{eqnarray*}
   A_1^k&=&\frac{(-1)^{k+1}}{8}\frac{(k+1)^2(2k+3)}{k+2},  \\
   A_2^k&=&\frac{(-1)^{k+1}}{8}\frac{(k+1)^2k^2}{k+2},  \\
   B^k&=&\frac{(-1)^{k+1}}{4}k^2\sqrt{\frac{(s-k-1)(s+k+2)}{2}}, \\
   C^k&=&\frac{(-1)^k}{16}(k+1)(s-k)(s+k+1).
\end{eqnarray*}
This Lagrangian differs from the one in ref.~\cite{Sing-Hagen}
only by the choice of normalization of the fields
\{$\Phi^s,\Phi^{s-2},\Phi^{s-3},\ldots,\Phi^0$\}.

\section{Electromagnetic Interaction of Spin 2 Particle}
\label{linear-sp2}

In this section we will investigate the electromagnetic interaction
of massive spin-2 particle in a linear approximation using the
constructive approach offered in ref.~\cite{Fronsdal-2,Zinoviev-1}.

Let us introduce the following notations. The superscripts will
denote a number of derivatives (both for the transformations and the
Lagrangians) while the subscripts will denote a number of fields for
the transformations and number of fields minus two for the
Lagrangians, i.e.:
$\delta_n^k\sim m^{1-(n+k)}\partial^k\Phi^n\Lambda$,
$\L_n^k\sim m^{2-(n+k)}\partial^k\Phi^{2+n}$.
In this notations a Lagrangian and transformations of any theory have
the general structure
\begin{eqnarray*}
    \L&=&\L^0_0+\L^1_0+\L^2_0+\L^0_1+\L^1_1+\ldots \\
    \delta&=&\delta^0_0+\delta^1_0+\delta^0_1+\de^1_1+\ldots
\end{eqnarray*}
In this, a variation of the Lagrangian has the form
\begin{eqnarray*}
  \delta
\L&=&\delta_0^0\L_0^0+\left(\delta_0^0\L_0^1+\delta_1^0\L_0^0\right)
  +\left(\delta_0^0\L_0^2+\delta_0^1\L_0^1\right)+\delta_0^1\L_0^2
   +\left(\delta_0^0\L_1^0 +\de_1^0\L_0^0\right) \\
&& {} +\left(\delta_0^0\L_1^1 +\delta_0^1\L_1^0
      +\de_1^0\L_0^1+\de_1^1\L_0^0\right) +\ldots
\end{eqnarray*}
It is easy to see that $\delta\L$ breaks into the sum of the
independent groups for which the sums of the superscripts and
subscripts are the same for every term of the group (in the linear
approximation we are going to consider the sum of subscripts which
less or equal to one). The different groups contain the terms with
different numbers of fields and/or derivatives, so the condition
$\delta\L=0$ means that each group should vanish independently and
this allows one to build an interaction by iterations on the number
of fields.

It is convenient to describe charged particles by complex fields.
Therefore, to begin with let us write the Lagrangian for complex
free spin-2 particle with mass $m$
\begin{eqnarray}
\label{Lfree2}
\non
    \L_0&=&\frac{1}{2}\partial_\mu\bar h_{\alpha\beta}
          \partial^\mu h^{\alpha\beta}
    -(\partial\bar h)_\mu(\partial h)^\mu
    +\frac{1}{2}\left((\partial\bar h)_\mu\partial^\mu h+h.c.\right)
    -\frac{1}{2}\partial_\mu\bar h\partial^\mu h  
\\\non &&{}
-\frac{1}{4}\bar B_{\mu\nu}B^{\mu\nu}
    +\frac{1}{2}\partial_\mu\bar\varphi\partial^\mu\varphi
    +\frac{m}{\sqrt{2}}\left(\bar h_{\mu\nu}\partial^\mu b^\nu
    -\bar h(\partial b)+h.c.\right) 
\\\non &&{}
+\frac{m\sqrt{3}}{2}(\bar b_\mu\partial^\mu\varphi+h.c)
    -\frac{m^2}{2}\left(\bar h_{\mu\nu}h^{\mu\nu}-\bar h h\right) 
\\ &&{}
+\frac{\sqrt{3}}{2\sqrt{2}}m^2\bar h\varphi+m^2\bar\varphi\varphi\ ,
\end{eqnarray}
where $h=g^{\mu\nu}h_{\mu\nu}$, 
$B_{\mu\nu}=\partial_\mu b_\nu-\partial_\nu b_\mu$ and the bar
denotes a complex conjugation. In this, the gauge transformations
have the form:
\begin{eqnarray}
    \delta_0h_{\mu\nu}&=&\partial_\mu\xi_\nu+\partial_\nu\xi_\mu
    +\frac{m}{\sqrt{2}}g_{\mu\nu}\eta, \\ \non
    \delta_0b_\mu &=&\partial_\mu\eta+m\sqrt{2}\xi_\mu,
\label{Vfree2} \\
    \delta_0\varphi &=&{}-m\sqrt{3}\eta. \non
\end{eqnarray}

In the notations given above the Lagrangian for free massive particle
has the structure $\L_0=\L_0^0+\L_0^1 +\L_0^2$, while the
transformations have the form $\delta_0=\delta_0^0+\delta_0^1$.

Let us switch on the electromagnetic interaction using the minimal
coupling prescription, i.e. we make the substitution $\partial_\mu
\rightarrow \D_\mu$, where $\D_\mu=\partial_\mu -iqA_\mu$ is the
covariant derivative. In this, we have to add $-\frac{1}{4}
(F_{\mu\nu})^2$ to~\reff{Lfree2}, where $F_{\mu\nu} = \partial_\mu
A_\nu - \partial_\nu A_\mu$. As usual, for the theory to be invariant
under the $U(1)_{em}$ transformations, the field $A_\mu$ must enter
through the covariant derivative or the $F_{\mu\nu}$ tensor only.
Such covariantization of the derivatives means the addition of the
terms of the form $\L_1^1,\L_1^0,\L_2^0$  to Lagrangian~\reff{Lfree2}
and the ones of the form  $\delta_1^0$ to
transformations~\reff{Vfree2}. But as a result of such substitutions,
due to the noncommutativity of the covariant derivatives, the
Lagrangian lost its invariance under the gauge transformations:
\begin{eqnarray*}
    \delta_1^0\L_0^2+\left(\delta_1^0\L_0^1+\delta_0^0\L_1^1\right)
&=& -iq\biggl\{-\sqrt{2}mF^{\mu\nu}\bar b_\mu\xi_\nu
    +\frac{1}{2}\partial_\mu F^{\mu\nu}\bar h\xi_\nu \\
&&  +\frac{3}{2}F^{\mu\nu}\D_\mu\bar h\xi_\nu
    -\partial_\mu F^{\mu\nu}\bar h_{\nu\alpha}\xi^\alpha
    -2F^{\mu\nu}\D_\mu\bar h_{\nu\alpha}\xi^\alpha \\
&&{}-F^{\mu\nu}(\D\bar h)_\mu\xi_\nu
    +\frac{1}{2}F^{\mu\nu}\D_\mu\bar b_\nu\eta\biggr\}+h.c.
\end{eqnarray*}

In order to recover the invariance of the Lagrangian let us add to
it and to the transformations new terms. We study the linear
approximation and it means that we add the linear terms to the
transformations and the cubic ones to the Lagrangian only. The number
of derivatives in the additional terms to the Lagrangian and to the
transformations must be consistent. For example, introducing the new
transformations of the form $\delta_1^n$, one has to add to the
Lagrangian the terms of the form $L_1^{n+1}$, because in calculation
of the variation they give a contribution of the same order%
\footnote{The order is defined by the number of derivatives} .  In
this, the transformations for $A_\mu$ is defined up to the terms
of the kind $\delta A_\mu\sim\partial_\mu (\ldots)$ because the
contribution of such terms to the variations vanishes due to the
$U(1)_{em}$ gauge invariance.

We will start with the minimal number of derivatives in the
additional terms and will increase this number until the gauge
invariance is recovered.

Let us add all possible linear terms without derivatives to the
transformations:
$$
\delta_1^0A_\mu=iq\left\{\alpha_1\bar h_{\mu\nu}\xi^\nu
       +\alpha_2\bar h\xi_\mu+\alpha_3\bar b_\mu\eta
       +\alpha_4\bar\varphi\xi_\mu\right\}+h.c.
$$
For the fields $h_{\mu\nu},b_\mu$ the appropriate terms have already
been added, because they are part of the covariant derivatives and
for the field $\varphi$ they are absent in this order.
A requirement of the closure of the algebra on the field $A_\mu$
imposes a nontrivial condition on the unknown coefficients:
$$
 \alpha_1+4\alpha_2+2\alpha_3-\sqrt{6}\alpha_4=0
$$
while on the fields $h_{\mu\nu},b_\mu,\varphi$ the closure of algebra
is trivial.

Correspondingly, the only possible in this order additional terms to
the Lagrangian have the form:
$$
\L_1^1=iqF_{\mu\nu}\left\{a_1\bar h_{\mu\alpha}h^\alpha{}_\nu
        +a_2\bar b_\mu b_\nu\right\}.
$$

The requirement of the gauge invariance:
\begin{eqnarray}
&&  \delta_1^0\L_0^1+\delta_0^0\L_1^1=0, \nonumber\\
&&  \delta_1^0\L_0^2+\delta_0^1\L_1^1=0
    \label{G-inv}
\end{eqnarray}
yields a non-homogeneous system of linear equations on the
coefficients in $\L_1^1$ and $\delta_1^0A_\mu$. But the system has no
solution, therefore, one needs add linear terms with one derivative
to the transformations and cubic terms with two derivatives to the
Lagrangian.

Let us write all such nontrivial additional terms with two
derivatives:
\begin{eqnarray*}
   \L_1^2&=&\frac{iq}{m}F^{\mu\nu}\Bigl\{
           b_1(\partial\bar h)_\mu b_\nu
          +b_2\bar h_{\mu\alpha}\partial^\alpha b_\nu
          +b_3\partial_\mu\bar h b_\nu+b_4\bar h \partial_\mu b_\nu
          +b_5\partial_\mu\bar b_\nu\varphi  \\
&&      {}+b_6\bar b_\nu\partial_\mu\varphi
          +b_7\partial_\mu\bar h_{\nu\alpha}b^\alpha
          +b_8\bar h_{\mu\alpha}\partial_\nu b^\alpha
          \Bigr\}+h.c.
\end{eqnarray*}
Correspondingly, all possible additional terms to the transformations
of the $A_\mu$ field with one derivative have the form:
\begin{eqnarray*}
    \delta_1^1 A_\mu&=&\frac{iq}{m}\Bigl\{
       s_{h1}(\partial\bar h)_\mu\eta
      +s_{h2}\bar h_{\mu\nu}\partial_\nu\eta
      +s_{h3}\partial_\mu\bar h\eta
      +s_{b1}(\partial\bar b)\xi_\mu
      +s_{b2}\bar b_\nu\partial^\nu\xi_\mu  \\
&&    +s_{b3}\partial^\nu\bar b_\mu\xi_\nu
      +s_{b4}\partial_\mu\bar b_\nu\xi^\nu
      +s_{\varphi}\bar\varphi\partial_\mu\eta
      \Bigr\}+h.c.
\end{eqnarray*}
while among the fields $h_{\mu\nu},b_\mu,\varphi$ only vector one has
such a term:
$$
  \delta_1^1 b_\mu=\frac{iq}{m}dF_{\mu\nu}\xi^\nu+h.c.
$$

Here in calculations we use the ordinary derivatives because in the
linear approximation they give the same result as the covariant
ones\footnote{The difference between the usual and the covariant
derivatives will be essential only in the next approximations.} .

The closure of the algebra on the electromagnetic field in this
order gives the nontrivial conditions for the unknown coefficients
in the transformations
\begin{eqnarray*}
&&   s_{b1}=s_{h1}=0, \qquad\\
&&   \alpha_1+\sqrt{2}(s_{b2}+s_{b3})=0,\qquad\\
&&   \sqrt{2}\alpha_2+s_{b5}=0,\\
&&   s_{b5}-2s_{h3}=0,\\
&&   s_{h3}+s_{b2}=0,\\
&&   s_{h2}-s_{b3}-s_{b4}=0.
\end{eqnarray*}

The gauge invariance requires that condition~\reff{G-inv} be
supplemented with:
$$
 \delta_0^1\L_1^2+\delta_1^1\L_0^2=0. 
$$
Besides, in this order new additional terms of the kind
$ \delta_1^1\L_{-1}^2$ appear, therefore, one has to modify the
second
equation in~\reff{G-inv}:
$$
 \de_0^1\L_1^1+\de_1^0\L_0^2+\de_1^1\L_0^1=0.
$$
One needs to take into account that not all the terms obtained from
the calculation of variation are independent. For example, its
necessary to use the Bianchi identity for the terms, which have the 
derivative on the $F_{\mu\nu}$ tensor
$$
\partial_\alpha F_{\mu\nu}+\partial_\mu F_{\nu\alpha}
   +\partial_\nu F_{\alpha\mu}=0. 
$$

All these conditions give a non-homogeneous system of linear
equations on the arbitrary coefficients. The solution of the system
of equations yields the result:
$$\L_{total}=\L_{free}+\L_{int},$$
where
\begin{eqnarray*}
    \L_{int}&=&iqF^{\mu\nu}\bar h_{\mu\alpha}h^\alpha{}_\nu
          +iqF^{\mu\nu}\bar b_\mu b_\nu \\
&& {}-\frac{iq}{m}F^{\mu\nu}\biggl(
      \frac{1}{\sqrt{2}}\bar h_{\mu\alpha}\partial^\alpha b_\nu
     -\frac{\alpha_2}{\sqrt{2}}\partial_\mu\bar h b_\nu
     -\frac{\alpha_2+\frac{1}{2}}{\sqrt{2}}\bar h\partial_\mu b_\nu
\\
&& {}+\frac{\alpha_4}{\sqrt{2}}\partial_\mu\bar b_\nu\varphi
     +\frac{\sqrt{\frac{3}{2}}+\alpha_4}{\sqrt{2}}
      \bar b_\nu\partial_\mu\varphi
     -\frac{\alpha_1+1}{\sqrt{2}}\partial_\mu
        \bar h_{\nu\alpha}b^\alpha \\
&& {}+\frac{\alpha_1}{\sqrt{2}}\bar h_{\mu\alpha}\partial_\nu
      b^{\alpha}  -h.c.\biggr).
\end{eqnarray*}
In this, we still have the relation for $\alpha_i$
$$
\alpha_1+4\alpha_2+2\alpha_3-\sqrt{6}\alpha_4=0.
$$
The transformations of the fields have the following form:
\begin{eqnarray*}
    \delta_1^1 b_\mu & =
&{}-\frac{iq}{m}\sqrt{2}F_{\mu\nu}\xi^\nu+h.c.\\
    \delta_1^1A_\mu & = &{}iq(\alpha_1\,\bar h_{\mu\nu}\xi^\nu
         +\alpha_2\bar h\xi_\mu+\alpha_3\bar b_\mu\eta
         +\alpha_4\bar\varphi\xi_\mu) \\
&&    {}+\frac{iq}{m}\biggl(
         \frac{\alpha_1+1}{\sqrt{2}}\bar h_{\mu\nu}\partial^\nu\eta
        -\frac{\alpha_2}{\sqrt{2}}\partial_\mu\bar h\eta
        -\frac{1+\alpha_1}{\sqrt{2}}\bar b_\nu\partial^\nu\xi_\mu \\
&&    {}+\frac{1}{\sqrt{2}}\partial^\nu\bar b_\mu\xi_\nu
        +\frac{\alpha_1}{\sqrt{2}}\partial_\mu\bar b_\nu\xi^\nu
        -\sqrt{2}\alpha_2\bar b_\mu(\partial\eta) \\
&& {}+\frac{\alpha_4+\sqrt{\frac{3}{2}}}{\sqrt{2}}
       \bar\varphi\partial_\mu \eta\biggr)+h.c.
\end{eqnarray*}

The presence of an arbitrariness in the Lagrangian and in the
transformations is related to a possibility to make a redefinition of
the field $A_\mu$, so that the physical content of theory remains
unchanged. This is the general situation for the theories which have
in an interaction the number of derivatives, which is equal to or
greater than the number of derivatives in a free Lagrangian. In this
order one has the three-parametric freedom in the definition of the
field $A_\mu$:
$$
 A_\mu\longrightarrow A_\mu+\frac{iq}{m}\left(
        c_1\bar h_{\mu\nu}b^\nu+c_2\bar h b_\mu+c_3\bar b_\mu
        \varphi\right)+h.c.
$$

Using this freedom one can choose the most convenient form for the 
Lagrangian
\begin{eqnarray}
\label{L2res}\non
   \L_{int}&=&iqF^{\mu\nu}\bar h_{\mu\alpha}h^\alpha{}_\nu
                +iqF^{\mu\nu}\bar b_\mu b_\nu  \\
&&  {}-\frac{iq}{m}F^{\mu\nu}\left(
       \frac{1}{\sqrt{2}}\bar h_\mu{}^\alpha B_{\alpha\nu}
      -\frac{1}{4\sqrt{2}}\bar h B_{\mu\nu}
      -\frac{\sqrt{3}}{4}B_{\mu\nu}\varphi-h.c.\right)
\end{eqnarray}
and the transformations
\begin{eqnarray*}
    \delta_1^1b_\mu&=&{}-\frac{iq}{m}\sqrt{2}F_{\mu\nu}\xi^\nu+h.c.
\\
    \delta_1^1A_\mu&=&{}-iq\left(
      \bar h_{\mu\nu}\xi^\nu+\bar b_\mu\eta
      +\sqrt{\frac{3}{2}}\bar\varphi\xi_\mu\right)
      -\frac{iq}{m\sqrt{2}}\bar B_{\mu\nu}\xi^\nu+h.c.
\end{eqnarray*}
The following choice of the parameters corresponds to such result:
$$
\alpha_1=-1, \quad \alpha_2=0, \quad \alpha_3=-1, \quad
\alpha_4=-\sqrt{\frac{3}{2}} .
$$

Further, one can proceed in one of the three ways. Firstly, one can
evade the introduction of any additional fields, but the most
probable situation in this case is that one will have to deal with
essentially non-linear theory. Secondly, one can introduce in the
system a finite number of additional fields and try to stop
iterations at some finite order. Thirdly, one can try to get a linear
theory introducing an infinite number of additional fields and using
the properties of infinite-dimensional algebras of the Kac-Moody
type.

Up to now only the third scenario has been realized in the theories
of Kaluza-Klein type. The simplest example is the reduction of
fifth-dimensional theory of gravity (see Ref.~\cite{Dolan}). A
linear approximation of this theory coincides with the result
obtained in this section. It is natural, because \reff{L2res}
contains a linear approximation of any theory with the number
of derivatives less than or equal to two. It is this universality of
the linear approximations that makes possible to consider this
important step independently of the scenario one could choose to
proceed further on.

The next quadratic approximation requires rather cumbersome
calculations even without introducing any additional fields.
Therefore in the next Section we will consider the simpler case
of homogeneous electromagnetic filed.

\section{Massive Spin-2 Particle in the Homogeneous Electromagnetic
Field}
\label{full-sp2}

Let us consider a massive spin-2 particle moving in a constant
homogeneous electromagnetic field.

Up to now we worked in the four-dimensional space. From this
point on we will use the flat Minkowski space of arbitrary dimension
$n$ with the signature of a metric $(+,--...)$. Let Latin indices
take the values $0,1,...,n-1$. For convenience we will not make
difference between upper and lower indices, while the summation over
repeated the indices will be understood, as usual, i.e.
$$
A_{k...}B_{k...}\equiv g^{kl}A_{k...}B_{l...} .
$$

This time we will choose the gauge invariant Lagrangian describing
free complex field with mass $m$ and spin 2 in the following form:
\begin{eqnarray}\label{l0}
{\cal L}_0 & = & \partial _m\bar h_{kl}\partial _mh_{kl}
\,-\,2\partial_kh_{kl}\partial _m\bar h_{lm}
\,+\,\left(\partial _kh_{kl}\partial _l\bar
h\,+\,h.c.\right)\,-\,\partial _k\bar h\partial _k h 
\non \\ &&{}
 +\,2\left(\partial _k\bar h_{kl}\partial _l\varphi 
\,-\,\partial _l\bar h\partial_l\varphi \,+\,h.c.\right)
\,-\,2\left(\partial _l\bar b_k\partial _lb_k\,-\,\partial
_lb_k\partial
_k\bar b_l\right) 
\non \\  &&{}
+\,2m\left(\partial _l\bar b_kh_{kl}\,-\,\partial _k\bar b_kh
\,+\,h.c.\right)\,-\,m^2\left(\bar h_{kl}h_{kl}\,-\,\bar hh\right) ,
\end{eqnarray}

The gauge transformations for this Lagrangian look like:
\begin{eqnarray}
\label{d0}
\delta h_{kl} & = & 2\partial _{(k}\xi _{l)} \non \\
\delta b_k & = & \partial _k\eta \,+\,m\xi _k \\ 
\delta \varphi & = & m\eta .\non
\end{eqnarray}

The Lagrangian has been chosen in a non canonical form (with the
off-diagonal kinetic terms) in order that the Goldstone part
(proportional to mass) for the field $h_{kl}$ was absent in
transformations~\reff{d0}. Lagrangian~\reff{l0} can be obtained
from~\reff{Lfree2} by redefining the second rank field
$$
 h_{kl}\to h_{kl}+\frac1{\sqrt{6}}g_{kl}\varphi
$$
and changing the normalizations of the fields and the gauge
parameters. Besides, unlike canonical form~\reff{Lfree2},
Lagrangian~\reff{l0} does not depend on dimensionality of the
space-time.

As before we will start with switching on the electromagnetic
interaction in \reff{l0} in the "minimal" way, i.e., we will
substitute the covariant derivatives instead of the ordinary ones.
For convenience, we put $m=1$ and change the definition of
electromagnetic field tensor by including the imaginary unit and the
charge $q$ in its definition, i.e. $iqF_{kl}\to F_{kl}$.

As usual, after switching on the minimal interactions we lose the
gauge invariance of Lagrangian~\reff{l0} under
transformations~\reff{d0}. The residual appears
\begin{eqnarray}
\label{невязка0} \non
\delta_{\bar0}\L_{\bar0}&=&4F_{kl}\D_l\bar h_{km}\xi _m
-2F_{lm}\D_k\bar h_{kl}\xi_m+3F_{kl}\D_k\bar h\xi _l
+6F_{km}\D_k\bar \varphi \xi _m 
\\ &&{}
-4F_{km}\bar b_k\xi _m-2F_{kl}\D_l\bar b_k\eta + h.c.\ ,
\end{eqnarray}
where the bar over an index denotes the replacement of partial
derivatives by the covariant ones. To compensate for this residual
one has to add new terms to the Lagrangian and the transformations.

Let us consider the linear approximation, i.e., we neglect the terms,
which are quadratic or higher in $F$. We will add to
transformations~\reff{d0} all possible terms containing up to one 
derivative\footnote{We have explicitly checked that if we restrict
ourselves by derivativeless transformations, we shall receive an
inconsistent system of equations in the next approximation.}
\begin{eqnarray}
\label{anz_d1}
\non
\delta_1h &\sim& F\partial\xi,
\\
\delta_1 b &\sim& F\partial\eta + F\xi,
\\\non
\delta_1\varphi &\sim& F\partial\xi.
\end{eqnarray}
For the case of homogeneous electromagnetic field the transformations
must not include more than one derivative, because in such case all
the derivatives act upon parameters of gauge transformations and if
the number of derivatives is more than one, the number of
physical degrees of freedom will change.

New transformations~\reff{anz_d1} give a contribution for the
variation in the following form\footnote{The calculations are rather
cumbersome and were made with the help of the "REDUCE" system.
Therefore, we will, as a rule, omit intermediate results and give
only their schematic representation.}
\begin{equation}
\label{вкладd1}
\delta_1\L_0=F\partial^3\bar h\xi+F\partial^3\bar b\eta
+F\partial^3\bar\varphi\xi+F\partial^2\bar h\eta+F\partial^2\bar b\xi
+F\partial\bar h\xi+h.c.
\end{equation}

The most general anzats for additional terms to the Lagrangian, which
give a contribution like~\reff{вкладd1} contains terms with, at most,
two derivatives
\begin{equation}
\label{anz_l1}
\L_1=F\partial \bar h \partial h+F\partial \bar h \partial\varphi
+F\partial \bar b\partial b 
+F\bar h\partial b + F\bar b\partial\varphi+F\bar h h+F\bar b b +
h.c.
\end{equation}

We will not consider the terms containing more than two derivatives.
To support this limitation, we could offer two arguments:
\begin{enumerate}
\item In the case of homogeneous electromagnetic field, $F$ is just a
constant matrix, that does not depend on the spatial coordinates.
Therefore, in all orders of the iterations the Lagrangian will remain
quadratic in the nontrivial fields $h_{kl},b_m, \varphi$
and the transformations will be always Abelian. Hence the Lagrangian
will be quasi-free and two derivatives are natural for it.
\item As we have already explained, there are no terms in the
transformations, which correspond to the terms in the Lagrangian
with more than two derivatives.
\end{enumerate}

From the requirement of the gauge invariance in the linear
approximation
$$
\delta_{\bar0}\L_{\bar0} + \delta_1\L_0 + \delta_0\L_1 = 0
$$
we obtain a nonhomogeneous system of linear equations for the
arbitrary coefficients in~\reff{anz_d1} and~\reff{anz_l1}\footnote{
As for all orders in $F$ the algebra is Abelian, its closure does not
give any additional conditions.} .

Solving the system of linear equations, we obtain the following
result for the transformations in the linear approximation
\begin{eqnarray}
\label{d1}\non
\delta_1h_{kl}&=&\alpha _1g_{kl}F_{mn}\partial _n\xi _m\ ,
\\
\delta_1b_k&=&\al_2F_{kl}\xi_l\ ,
\\\non
\delta_1\varphi&=&\alpha _3F_{kl}\partial _l\xi _k\ .
\end{eqnarray}
while for the Lagrangian we get:
\begin{eqnarray*}
{\cal L}_1&=&
\left(\al_1\left(n-1\right)-\al_2+1\right)
F_{kl}\bar h_{km}h_{lm}+4F_{kl}\bar b_kb_l
\\ &&{} 
+\left\{\left(\al_1\left(n-1\right)-\al_2-3\right)
F_{kl}\partial _l\bar b_mh_{km}
\right.\\ &&{}
+\left(\al_1\left(n-1\right)+\al_2+3\right)F_{kl}\partial _m\bar
b_lh_{km} 
\\ &&\left.{\!}
+\left(\al_1(n-1)+2\al_2+3\right)F_{kl}\partial _l\bar
b_kh_{mm}-6F_{kl}\partial _lb_k\bar \varphi +h.c.\right\}
\\ &&{} 
+\left(\alpha _1\left(n-2\right)+2\alpha_3\right)
\left(F_{ln}\partial _mh_{kl}\partial _m\bar h_{kn}-
F_{ln}\partial _kh_{kl}\partial _m\bar h_{mn}\right) 
\\ &&{} 
+\left\{\left(\alpha _1\left(n-2\right)+2\alpha_3\right)
\left(F_{mn}\partial _kh_{kl}\partial _n\bar
h_{lm}+F_{ln}\partial _kh_{kl}\partial _n\bar h\right)
\right.\\ && \left.{\!}
-2\al_1\left(n-1\right)F_{km}\partial _k\bar \varphi \partial
_lh_{lm}
-\left(\al_1\left(n-1\right) + 3\al_2+3\right)
F_{kl}\partial _l\bar b_k\partial _mb_m
\right.\\ &&\left.{\!}
+h.c.\vphantom{\sqrt{F}}\right\}
-\left(\al_1\left(n-1\right) + 3\al_2+3\right)
F_{km}\partial_lb_k\partial _l\bar b_m.
\end{eqnarray*}
In this, we have used a two-parametric freedom related to a
possibility of field redefinitions:
\begin{eqnarray*}
 h_{kl}&\to&h_{kl}+s_hF_{(k|m}h_{|l)m},
\\
b_k&\to&b_k+s_bF_{kl}b_l.
\end{eqnarray*}

Now, going to a quadratic approximation, we get the residual,
containing the terms quadratic in $F$, which appears from the
$\de_{\bar1}\L_{\bar0}+\de_{\bar0}\L_{\bar1}+\de_1\L_1$.
To compensate the residual we proceed in the same manner as in the
linear approximation i.e. we add the terms of the form
\begin{equation}
\label{anz_d2}
\de_2\Phi \sim FF\partial\Lambda + FF\Lambda
\end{equation}
to the transformations, where $\Phi\sim\{h_{kl},b_k,\varphi\}$, and
$\Lambda\sim\{\xi_k,\eta\}$.
At the same time, we add all the possible terms of the form
\begin{equation}
\label{anz_l2}
\L_2= FF\partial\bar\Phi\partial\Phi+
FF\partial\bar\Phi\Phi+FF\bar\Phi\Phi
\end{equation}
to the Lagrangian.
Imposing the condition of the gauge invariance in the quadratic
approximation, i.e.
$$
\de_{\bar1}\L_{\bar0}+\de_{\bar0}\L_{\bar1}
+\de_1\L_1+\de_2\L_0+\de_0\L_2=0\ ,
$$
one obtains a system of quadratic equations for the coefficients in~%
\reff{anz_d2}, \reff{d1} and~\reff{anz_l2}. Solving the system we get
an answer to the given order.

Note, that in this order\footnote{This holds for any even order.} \ 
one can exclude all the terms in the transformations proportional to
$F^2$ using the freedom in the field redefinition
$\Phi\to\Phi+FF\Phi$.

In some sense the quadratic approximation is crucial, because if the 
coefficients $\al_i$ in~\reff{d1} are not equal to zero, then
one can exclude all the transformations in the higher orders using
the freedom in the field redefinition for every order as well as an
arbitrariness in the definition of the $F$ tensor $F\to
F+F^3+F^6+\ldots$.

Let us try to stop the iterations. For this we will decrease the
number of derivatives for each next order of the iteration. That is
we are adding the terms of the kind
\begin{eqnarray*}
\L_3&\sim& FFF\bar\Phi\Phi+FFF\bar\Phi\Phi\ ,
\\
\L_4&\sim& FFFF\bar\Phi\Phi\ .
\end{eqnarray*}

Requiring the gauge invariance for all orders, we obtain a system of 
non-homogeneous algebraic equations of the fourth degree. Solving
this system, we get the final answer for the Lagrangian describing a
charged massive spin-2 particle moving in the constant homogeneous
electromagnetic field
\begin{equation}
\label{lfull}
\L_{full}=\L_{\bar0}+\L_{\bar1}+\L_{\bar2}+\L_{\bar3}+\L_{\bar4}\ ,
\end{equation}
where
\begin{eqnarray*}
\L_{\bar1}&\!=\!&\frac54F_{kl}\bar h_{km}h_{lm}+2F_{kl}\bar b_kb_l
-\frac32F_{kl}\D_l\bar b_mh_{km}
+\frac32F_{kl}\D_m\bar b_lh_{km} 
\\ &&{} 
+6F_{kl}\D_l\bar b_k\varphi +\frac34F_{km}\D _lb_k\D
_l\bar b_m+\frac32F_{kl}\D _l\bar b_k\D_mb_m+h.c. 
\\
\L_{\bar2}&\!=\!&\frac94F_{kl}F_{mn}\bar
h_{ln}h_{km}+\frac32\left(\frac38F_{kl}F_{km}
\bar h_{lm}h-F_{kl}F_{km}\bar h_{lm}\varphi +h.c.\right) 
\\ &&{} 
-3\left(F_{kl}F_{km}\bar b_mb_l+\frac14F_{kl}^2\bar
b_mb_m-F_{kl}^2\bar 
\varphi\varphi \right)-\frac94\left(F_{km}F_{ln}\D_l\bar b_kh_{mn}
\right. \\ && \left.{\!}
+F_{km}F_{mn}\D _l\bar b_kh_{ln}+\frac12F_{lm}F_{ln}\D_k\bar
b_kh_{mn}+\frac12F_{km}F_{lm}\D_l\bar b_kh+h.c.\right) 
\\ &&{} 
-6\!\left(F_{kl}F_{lm}\D_k\bar \varphi b_m+h.c.\right)
+\frac9{40}\!\left(\frac32F_{np}^2\D
_k\bar h_{kl}\D_mh_{lm}-\frac34F_{np}^2\D_m\bar h_{kl}\D
_mh_{kl} 
\right.\\ &&{} 
-\frac32F_{np}^2\D_mh_{lm}\D_l\bar h+\frac34F_{np}^2\D_l
\bar h\D_lh+\frac14F_{np}F_{mp}\D_mh_{kl}\D_n\bar h_{kl} 
\\ &&{} 
-\frac12F_{np}F_{mp}\D _lh\D _l\bar h_{mn}+\frac12F_{np}F_{mp}
\D_k\bar h_{mn}\D _lh_{kl}-F_{np}F_{mp}\D _lh_{kl}
\D_n\bar h_{km} 
\\ &&{} 
+\frac12F_{np}F_{lp}\D _m\bar h_{kl}\D _mh_{kn}+F_{np}F_{lp}
\D_k\bar h_{kl}\D _nh-\frac12F_{np}F_{lp}\D _k\bar h_{kl}
\D_mh_{mn} 
\\ &&{} 
-\frac14F_{np}F_{lp}\D _l\bar h\D _nh+\frac92F_{np}F_{lm}
\D _m\bar h_{kl}\D _ph_{kn}-9F_{np}F_{lm}\D_k\bar h_{kl}
\D _ph_{mn}
\\ && \left.{\!}
+\frac92F_{np}F_{lm}\D _l\bar h\D _ph_{mn}-\frac94F_{lp}F_{kn}
\D_m\bar h_{kl}\D _mh_{np}+h.c.\right) 
\\ &&{} 
+\frac94\left(F_{ln}F_{mn}\D _kh_{kl}\D_m\bar \varphi
-F_{lm}F_{kn}\D_mh_{kl}\D_n\bar \varphi
-\frac12F_{kn}F_{ln}\D _mh_{kl}\D_m\bar \varphi  
\right.\\ && \left.{\!}
-\frac12F_{km}F_{lm}\D _lh\D_k\bar \varphi
+h.c.\right)-\frac92F_{kl}F_{mn}\D_lb_k\D_n\bar
b_m-6F_{km}F_{lm}\D_l\bar \varphi \D _k\varphi ,
\\
\L_{\bar3}&=&\frac{27}{80}F_{kl}^2F_{mn}\bar
h_{np}h_{mp}-\frac{153}{40}F_{kl}F_{km}
F_{np}\bar h_{mp}h_{ln}+\frac9{10}F_{kl}F_{km}F_{ln}\bar h_{mp}h_{np} 
\\ &&{} 
+\frac94\left(\frac32F_{kl}F_{mp}F_{mn}\D _lb_k\bar
h_{np}+F_{lm}^2F_{kn}\D
_k\varphi \bar b_n+h.c.\right) ,
\\
\L_{\bar4}&=&{}-\frac{27}{16} \left(F_{kl}^{2} F_{mq} F_{mn} h_{nq}
\bar{\varphi }
+h.c.\right)+\frac98 F_{kl}^{2} F_{mn}^{2} \bar{\varphi } \varphi
\\ &&
 +\frac{81}{32} F_{kl} F_{km} F_{np} F_{nq} \bar{h}_{lm} h_{pq}.
\end{eqnarray*}
In this, only field $b_k$ has the nontrivial transformation
\begin{equation}
\label{Vfin}
\de_1 b_k={}-\frac32F_{kl}\xi_l.
\end{equation}
It is easy to see that the result obtained does not depend on the
dimensionality of the space-time\footnote{We have to remark that one
should consider the case of two-dimensional space-time separately.} .
This is a consequence of our choice of noncanonical form for the
free Lagrangian and the absence of the term proportional to
the metric tensor in the transformations.

As it was already mentioned in the Introduction a similar problem in
a context of the bosonic string theory was discussed in
paper~\cite{Argyres}. Comparing the result
obtained there with lagrangian~\reff{lfull} and
transformations\footnote{One must replace the ordinary derivatives by
the covariant ones.} \ \ \reff{d0} and~\reff{Vfin}, one can make the
conclusion that both models have the same structure. That is, the
transformations of fields are linear for electromagnetic field $F$,
the number of the derivatives in the Lagrangians does not exceed two
and a maximal order in $F$ equals four. But since the authors of
ref.~\cite{Argyres} started from the bosonic string they obtained the
gauge invariant description of massive spin-2 field moving in the
homogeneous electromagnetic field in the form, which is valid only
for the 26-dimensional space-time.

\section*{Conclusion}

Thus, in this work the gauge invariant description of free massive
fields with arbitrary integer spins was constructed. Basing on such
description, one can investigate consistent theories of interactions
of massive particles with high spins. As an example of such
constructive approach, we considered the electromagnetic interaction
of massive spin-2 field and obtained: a) the linear approximation
with minimal number of the derivatives in the case of the arbitrary
electromagnetic field; b) the full answer for the homogeneous field
in the space-time of any dimensionality.

Later, using the offered method we are planning to consider the
electromagnetic interaction of a massive spin-3 field and also to
investigate possibilities to go beyond the linear approximation for
the spin-2 field in a case of the arbitrary electromagnetic field.

\vspace{0.3in}
{
\centerline{\large \bf Acknowledgments}
}
\vspace{0.2in}

Work supported by Russian Foundation for Fundamental Research 
grant 95-02-06312.

%\newpage

\end{document}